\documentclass{ifacconf}

\usepackage{graphicx}      % include this line if your document contains figures
\usepackage{natbib}        % required for bibliography

\begin{document}
\begin{frontmatter}

\title{Industrial AI Robustness Card for Time Series Models\thanksref{ifacaccepted}}

\thanks[ifacaccepted]{\copyright{} 2026 the authors. This work has been accepted to IFAC for publication under a Creative Commons Licence CC-BY-NC-ND.}

% \thanks[footnoteinfo]{Sponsor and financial support acknowledgment
% goes here. Paper titles should be written in uppercase and lowercase
% letters, not all uppercase.}

\author[First]{Alexander Windmann} 
\author[Second]{Benedikt Stratmann} 
\author[Third]{Mariya Lyashenko}
\author[Fourth]{Oliver Niggemann}

\address[First]{Institute of Artificial Intelligence, Helmut Schmidt University Hamburg, Germany (e-mail: alexander.windmann@hsu-hh.de).}
\address[Second]{Fraunhofer Institute of Optronics, System Technologies and Image Exploitation (IOSB),
    Karlsruhe, Germany
  (e-mail: benedikt.stratmann@iosb.fraunhofer.de)}
\address[Third]{Siemens AG, Digital Industries, Process Automation,
% Oestliche Rheinbrueckenstr. 50, 76187 
Karlsruhe, Germany (e-mail: mariya.lyashenko@siemens.com)}
\address[Fourth]{Institute of Artificial Intelligence, Helmut Schmidt University Hamburg, Germany (e-mail: oliver.niggemann@hsu-hh.de).}

\begin{abstract}                % Abstract of 50--100 words
Industrial AI practitioners face vague robustness requirements in emerging regulations and standards but lack concrete, implementation-ready protocols.
This paper introduces the Industrial AI Robustness Card for Time Series (IARC-TS), a lightweight protocol for documenting and evaluating industrial time series models.
IARC-TS specifies required fields and an empirical measurement and reporting protocol that combines drift and operational domain monitoring, uncertainty quantification, and stress tests, and maps these to selected EU AI Act documentation, testing, and monitoring obligations.
A biopharmaceutical soft sensor case study illustrates how IARC-TS supports reproducible robustness evidence and defines monitoring triggers.
\end{abstract}


\end{frontmatter}
%===============================================================================

\section{Introduction}

Empirical studies such as \cite{windmann_artificial_2024} identify a lack of trustworthiness as a central barrier to Artificial Intelligence (AI) adoption: AI models are often not robust, while assurance and explainability techniques remain difficult to apply to complex neural networks.
For industrial time series models, nominal test accuracy alone is weak evidence: sensor drift, missing channels, changing sampling rates, and operating-mode shifts can invalidate a model after deployment.
Furthermore, regulators and standardization bodies demand stronger requirements for monitoring and documenting AI systems.
For example, the \cite{RegulationEU20242024} requires risk management, technical documentation, and post-market monitoring for high-risk systems.
Frameworks and guidance such as the NIST AI Risk Management Framework by \cite{tabassiArtificialIntelligenceRisk2023} and \cite{ISOIEC238942023} call for systematic risk evaluation, monitoring, and documentation of trustworthiness properties such as robustness, but they do not prescribe concrete robustness metrics or test scenarios.
More technical robustness guidance, such as \cite{ISOIEC2402922023}, focuses on selecting and applying formal methods to prove robustness properties of neural networks.
This supports bounded-property analysis but does not by itself replace empirical stress tests for typical industrial deployments \citep{perez-cerrolazaArtificialIntelligenceSafetyCritical2024}.
Guidelines such as \cite{easaEASAArtificialIntelligence2024} provide more detailed evaluation recommendations but focus on domains such as aviation and are therefore only partially relevant for industrial time series.
In parallel, Model Cards introduced by \cite{mitchellModelCardsModel2019} and the IBM FactSheet by \cite{arnoldFactSheetsIncreasingTrust2019} offer compact templates to describe model scope, intended use, and limitations, but generally focus on \emph{what} to document rather than on \emph{how} to measure and report robustness or uncertainty.

Industrial AI practitioners therefore face a mismatch.
Emerging regulations and standards describe high-level robustness and monitoring obligations, but projects still need a concrete, lightweight protocol that turns these obligations into empirical tests, metrics, and documentation for industrial time series, see also \cite{diaz-rodriguezConnectingDotsTrustworthy2023}.
To address this gap, we propose the \emph{Industrial AI Robustness Card for Time Series} (IARC-TS), a card-style protocol for empirical robustness evaluation and monitoring of industrial time series models.
Our main contributions are:
\begin{itemize}
    \item A minimal schema for documenting and evaluating industrial time series models.
    \item A measurement and reporting protocol for uncertainty quantification and robustness evaluation.
    \item A compact mapping from IARC-TS fields to selected documentation, testing, and monitoring obligations of the EU AI Act.
    \item A soft-sensor case study with released code and a human- and machine-readable IARC-TS template.\footnote{https://github.com/awindmann/Industrial-AI-Robustness-Card}
\end{itemize}

\section{Background and Related Work}

Work on trustworthy AI emphasizes lifecycle assurance for data-centric systems.
We focus on industrial time series, where sensor streams support monitoring, prognostics, quality assurance, and diagnosis in cyber-physical systems \citep{niggemannMachineLearningCyberPhysical2023}.
\cite{ashmoreAssuringMachineLearning2021} and \cite{lavinTechnologyReadinessLevels2022} stress evidence across development, deployment, and monitoring, while \cite{perez-cerrolazaArtificialIntelligenceSafetyCritical2024} note that robustness and validation remain open challenges for industrial systems.
For chemical engineering, \cite{dobbelaereMachineLearningChemical2021} show that machine learning offers opportunities but requires vigilance about interpretability limits and overfitting, matching broader Industrial AI analyses in which robustness and interpretability remain adoption barriers \citep{windmann_artificial_2024}.

Documentation templates address part of this problem.
Model Cards by \cite{mitchellModelCardsModel2019}, IBM FactSheets by \cite{arnoldFactSheetsIncreasingTrust2019}, and EU-oriented model-reporting work by \cite{brajovicMergingEURegulationModel2023} standardize fields for intended use, performance, and limitations.
They leave substantial freedom in how robustness tests, calibration checks, and monitoring evidence are generated.
Dataset specifications add provenance and preprocessing detail \citep{hutchinsonAccountabilityMachineLearning2021}, but do not define an operational robustness procedure for deployed time series models.

Technical robustness guidance also remains too broad for the present use case.
\cite{ISOIECTR2021} and \cite{DINSPEC9200122020} describe robustness assessment processes, while \cite{easaEASAArtificialIntelligence2024} gives aviation AI assurance guidance, but for the model evidence targeted here these sources do not define a compact empirical protocol combining operational-domain checks, realistic stress scenarios, uncertainty metrics, and card-style reporting.
Similarly, \cite{ISOIEC420012023} specifies AI management-system requirements and \cite{ISOIEC238942023} gives AI risk-management guidance without selecting concrete tests for soft sensors or monitoring models.
Measuring trustworthiness remains difficult because many traditional software-safety and formal-verification methods do not apply cleanly to complex data-driven models \citep{perez-cerrolazaArtificialIntelligenceSafetyCritical2024}.
IARC-TS therefore complements formal analyses with realistic stress tests and natural perturbations, drawing on severity-controlled sensor-fault benchmarking for forecasting and uncertainty-quantification surveys \citep{windmannQuantifyingRobustnessBenchmarking2025,gawlikowski2022surveyuncertaintydeepneural}.
A compact, implementation-oriented card protocol translating trustworthy AI frameworks into robustness, calibration, and monitoring evidence for industrial time series models is still missing, consistent with \cite{diaz-rodriguezConnectingDotsTrustworthy2023}.
IARC-TS turns these elements into an implementation-oriented reporting protocol.

\section{Industrial AI Robustness Card for Time Series}
\label{sec:IARC}
IARC-TS documents model-level evidence rather than the complete, integrated AI system.
This scope provides model-side evidence that can feed selected high-risk AI documentation, testing, record-keeping, and post-market monitoring obligations under the \cite{RegulationEU20242024}, including Arts.~9--13, 15, and~72 and Annex~IV, but it is not a conformity-assessment artifact by itself.
Plant-level risk management, hardware, operator procedures, and cybersecurity evidence remain part of the surrounding assurance case.
This section describes the fields recorded by the card and the rationale for each group.
A concise time series measurement protocol that explains how to generate the evidence for these fields is provided in Section~\ref{sec:protocol}.

\subsection[General Information]{General Information\\(\cite{RegulationEU20242024} Art.~11, 13 (3), Annex~IV)}
The general information section summarizes the AI model and the associated dataset.
It records model and dataset name, version, date, provider, deployment context, and stable identifiers for later monitoring.
Tracking versions is important because the same card has to support lifecycle comparison after retraining or deployment updates \citep{tabassiArtificialIntelligenceRisk2023}.

\subsection{Intended Use (\cite{RegulationEU20242024} Art.~13)}
The intended use section states approved uses of the model and the decisions it may support, following model-card practice \citep{mitchellModelCardsModel2019,arnoldFactSheetsIncreasingTrust2019}.
It also states explicitly excluded uses, so that misuse can be identified during review.

\subsection{Data (\cite{RegulationEU20242024} Art.~10(2--4), Annex~IV(2)(d))}
For reliable deployment, the card documents dataset provenance, preprocessing steps, data quality, and data drift.
It also describes the distribution expected during operation and the stress-test scenarios that probe edge cases.
To this end, the card contains an Operational Design Domain (ODD), together with the part of that domain supported by training data.
This supports monitoring because drift out of the training-supported ODD can be detected early, while gaps between the engineering ODD and the training-supported region help define realistic stress-test scenarios.
The data section therefore links routine data documentation to scenario design rather than treating them as separate reporting tasks.

\subsection[Evaluation]{Evaluation\\(\cite{RegulationEU20242024} Arts.~9(6--8), 13(3)(b), and~15(1--4))}
AI models pose risks that differ from traditional software, because data drift can affect functionality and trustworthiness in ways that are hard to understand \citep{tabassiArtificialIntelligenceRisk2023}.
IARC-TS addresses this limitation by emphasizing multiple task-appropriate key performance indicators (KPIs), explicit uncertainty quantification, realistic robustness evaluation, and monitoring actions.
By evaluating performance on multiple KPIs with predefined acceptance thresholds, the card gives a more complete view of model capability than clean accuracy alone.
Uncertainty metrics help identify overconfident or unreliable predictions early.
The robustness field focuses on realistic stress-test scenarios rather than artificial adversarial attacks or purely formal stability guarantees, which often do not show which real-world perturbations can be tolerated \citep{ashmoreAssuringMachineLearning2021}.
A card is complete only when each reported metric is tied to a dataset split, model version, scenario definition, acceptance threshold, and monitoring action.

\subsection{Limitations (\cite{RegulationEU20242024} Art.~13(3)(b), Annex~IV(3))}
The limitations section records known limitations of both the AI model and the data \citep{brajovicMergingEURegulationModel2023}.
Examples include regimes where the model is not expected to perform well, reliance on simulated edge cases that have not yet occurred in practice, and restricted ODD coverage.
By making such limits explicit, IARC-TS supports realistic expectations and more transparent risk assessment.

\section{Measurement and Reporting Protocol}
\label{sec:protocol}

This section provides a time series protocol for generating IARC-TS evidence, informed by best-practice guidance on reliable and reproducible results \citep{vranjesDesignPrinciplesFalsifiable2024}.

\subsection{Data}

\subsubsection{Data quality}
Before modeling, compute missingness, summary statistics, and drift diagnostics for key variables before and after preprocessing.
Assess drift across time windows, batches, or operating regimes and estimate simple trends for critical features \citep{WebbCharacterizingconceptdrift2016}.
Where possible, retain raw and preprocessed diagnostics so the card records whether preprocessing hides or removes data-quality problems.

\subsubsection{Reproducibility controls}
Fix random seeds for splitting, initialization, and training procedures.
Record code and environment identifiers such as repository commits and library versions.
Assign dataset version identifiers that link raw data to preprocessing configurations and splits.
Store these identifiers with reported metrics and plots, and for retrained models preserve the previous card so clean-performance gains can be checked against scenario-robustness losses.

\subsubsection{Data splits}
Construct training, validation, and test sets with temporal or group-aware separation.
For sequential time series, use chronological splits instead of random cross-validation to avoid look-ahead leakage.
Alternatively, when dependencies across batches matter, split by these groups and avoid overlaps across splits.
Apply purge windows between adjacent splits to reduce contamination from temporal autocorrelation.

\subsubsection{ODD definition}
Following the ODD concept in \cite{easaEASAArtificialIntelligence2024}, capture asset and site characteristics, operating modes, and relevant environmental ranges.
For card-level reporting, operationalize this for a monitored state or feature vector $x_t\in\mathrm{R}^d$ as
\[
\begin{array}{l}
\mathcal{E}=\{x\in\mathrm{R}^d: l_j\le x_j\le u_j,\ j=1,\ldots,d\},\\[1mm]
\mathcal{O}_{\alpha}=\{x\in\mathcal{E}: \hat f_{\mathrm{tr}}(x)\ge\lambda_{\alpha}\}.
\end{array}
\]
where $\mathcal{E}$ is the engineering operating envelope, $[l_j,u_j]$ encode engineering limits, $\hat f_{\mathrm{tr}}$ is a kernel density estimate (KDE) of the training density, and $\lambda_{\alpha}$ is the density-level cutoff chosen so that approximately $1-\alpha$ of the training reference data lie in $\mathcal{O}_{\alpha}$.
Thus, $\mathcal{E}$ is the engineering ODD, and $\mathcal{O}_{\alpha}\subseteq\mathcal{E}$ is its training-supported region.
Points outside $\mathcal{E}$ leave the declared engineering envelope.
Points in $\mathcal{E}\setminus\mathcal{O}_{\alpha}$ remain inside the engineering ODD but have low support under the training reference distribution or reflect drift away from it.
Such cases should prompt operators to review uncertainty estimates and task performance, tighten monitoring, and target validation or stress tests.

\subsubsection{Scenario catalog}
Define relevant stress test scenarios aligned with the ODD. 
Prefer real slices such as batches, machines, or operating regimes.
When near-fail data are scarce, simulate plausible scenarios, for example sensor faults or noise patterns, see \cite{windmannQuantifyingRobustnessBenchmarking2025}.
Record the perturbed variable, target handling, severity levels, and operational rationale.

\subsubsection{Distributional diagnostics}
Quantitatively show how well the data represent the ODD and where test scenarios deviate.
Produce KDE plots for key features and compute distance measures between training data and each test scenario, for example using Kolmogorov–Smirnov tests or using the Wasserstein distance.
Rank features or scenarios by deviation from the baseline.

\subsection{Uncertainty quantification}
Select an uncertainty quantification mechanism appropriate for the model class and deployment constraints, see \cite{gawlikowski2022surveyuncertaintydeepneural}.
Examples include deep ensembles, Monte Carlo dropout, quantile regression, conformal prediction, or parametric predictive distributions.
If post-hoc calibration is used, apply it on a separate calibration set.
For interval forecasts, IARC-TS requires two reported uncertainty diagnostics: prediction interval coverage and the weighted interval score (WIS).
For a target level $1-\alpha$, report prediction interval coverage \citep{gawlikowski2022surveyuncertaintydeepneural} as
\[
\hat c_{\alpha}=n^{-1}\sum_{i=1}^{n}\mathbf{1}\{y_i\in[l_i^{\alpha},u_i^{\alpha}]\}.
\]
Here, $n$ is the number of evaluated cases, $y_i$ is the realized value, and $[l_i^{\alpha},u_i^{\alpha}]$ is the central prediction interval produced for the input of case $i$.
For an observation $y$ and interval $[l,u]$, define the per-level interval score \citep{gneitingStrictlyProperScoring2007}
\[
\mathrm{IS}_{\alpha}=(u-l)+2\alpha^{-1}(l-y)_+ +2\alpha^{-1}(y-u)_+
\]
where $(z)_+=\max(z,0)$.
For $K$ central intervals with miscoverage levels $\alpha_1,\ldots,\alpha_K$ and median $m$, the weighted interval score is \citep{bracherEvaluatingEpidemicForecasts2021}
\[
\mathrm{WIS}=\frac{1}{K+1/2}\left(\frac{1}{2}|y-m|+\sum_{k=1}^{K}\frac{\alpha_k}{2}\mathrm{IS}_{\alpha_k}\right).
\]
Report coverage and mean WIS over evaluated cases on the regular test data and each stress-test scenario.
For time series classification variants, replace coverage and WIS with negative log-loss or Brier score and a calibration diagnostic.

\subsection{Robustness}
Use the scenarios from the scenario catalog to test the AI model under realistic edge conditions.
For each scenario $s$, compute all KPIs and uncertainty metrics and compare them to the clean test baseline.
Let $\mathcal{S}$ be the scenario catalog and let $e_c>0$ and $e_s>0$ denote the clean and scenario values of a lower-is-better KPI $e$.
Define
\[
r_s(e)=\min(1,e_c/e_s),\qquad R(e)=|\mathcal{S}|^{-1}\sum_{s\in\mathcal{S}}r_s(e).
\]
Use $R(e)$ as an aggregated robustness score when a single KPI summary is needed.
For higher-is-better KPIs, use $r_s(e)=\min(1,e_s/e_c)$.
When simulated perturbations are used, vary severity and show severity-performance curves or aggregate robustness scores \citep{windmannQuantifyingRobustnessBenchmarking2025}.
Identify weakest scenarios and relate them to ODD factors to inform mitigation, monitoring, or risk acceptance.

\subsection{Monitoring and Actions}
Deployments should reuse diagnostics over fixed review windows and predefine warning and action thresholds for ODD and drift distance, KPI degradation, coverage error $|\hat c_\alpha-(1-\alpha)|$, and robustness-score changes.
Warnings trigger review, recalibration, or retraining, while action thresholds trigger rollback, fallback operation, or human intervention.
For closed-loop use, the plant assurance case must define the fallback controller and stability argument.
IARC-TS records model-side trigger logic, metric values, model and dataset versions, the scenario catalog, and the responsible operator or service per event.

\begin{figure*}[!t]
  \centering
\includegraphics[width=0.90\textwidth]{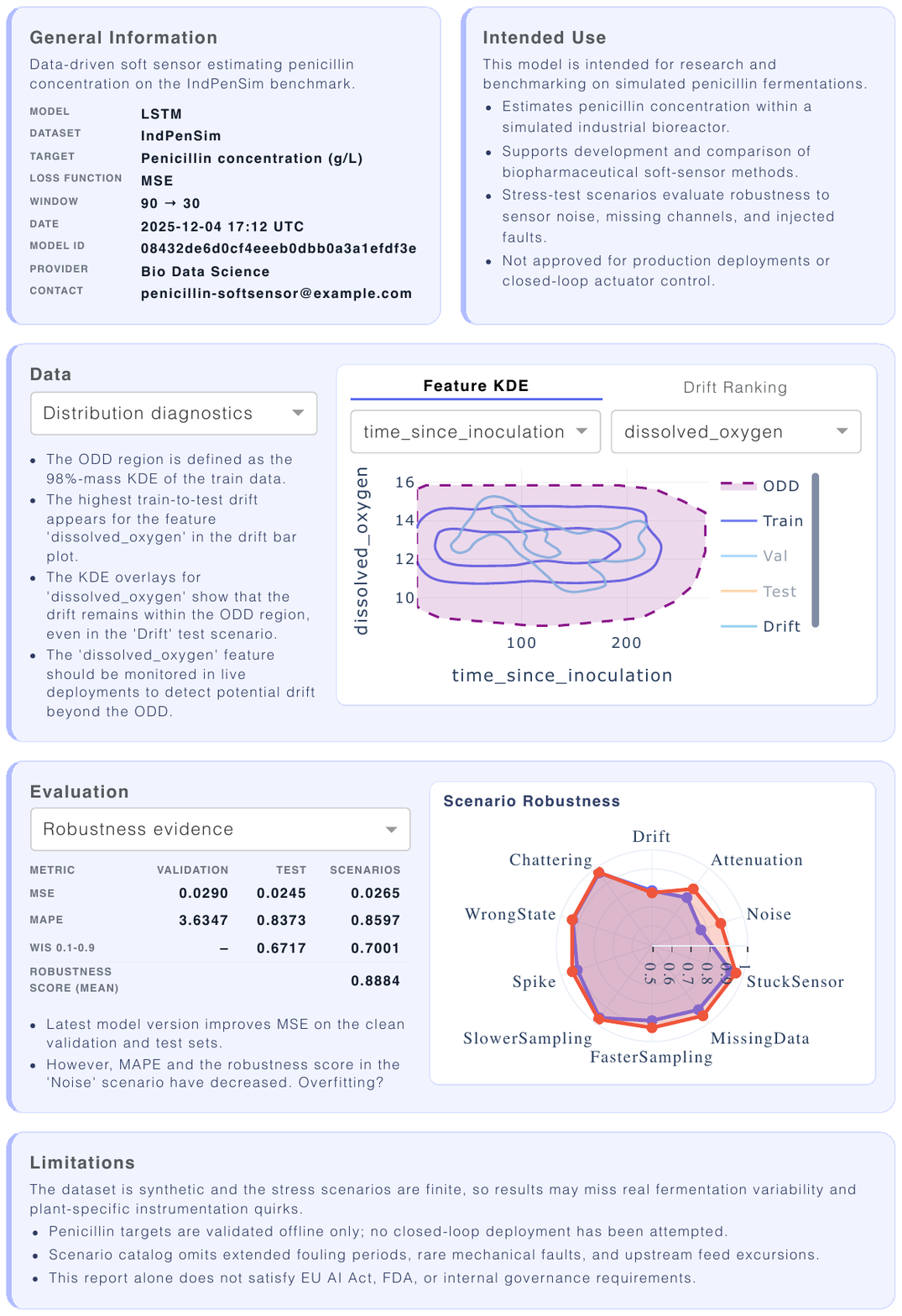}
  \caption{Example IARC-TS for IndPenSim. Some fields and plots are hidden due to space constraints.}
  \label{fig:card}
\end{figure*}

\section{Case Study}

\subsection{System and Task}
Validated soft sensors can reduce offline assays that delay biopharmaceutical quality decisions.
Our case study uses the IndPenSim industrial-scale fed-batch fermentation simulator by \cite{GOLDRICK2019106471}.
The LSTM soft sensor forecasts penicillin concentration from a 90-step input window to a 30-step horizon using mean squared error (MSE) loss.

Soft sensors that inform product-quality operate in a regulated manufacturing context.
Good Manufacturing Practice (GMP) guidance on AI \citep{GMPAIRegulations} points to documentation, validation, performance-metric, data-quality, monitoring, and human-review expectations.
If such a system is deployed in a high-risk use case under the \cite{RegulationEU20242024}, formal AI-system documentation obligations would also apply, with IARC-TS as one robustness evidence module.

\subsection{Results}

Figure~\ref{fig:card} shows the generated IARC-TS, where the ODD is a 98\%-mass KDE over the training distribution and drift diagnostics rank dissolved oxygen as the largest train-to-test shift.
The artifacts trace each evaluated run to its scenario catalog, metrics, and forecast-sample JSON files with clean and perturbed inputs, scenario names, and severities.
The catalog covers drift, attenuation, noise, stuck sensors, missing data, sampling changes, spikes, wrong-state substitution, and chattering.
Across scenarios, MSE changes from 0.0245 on the clean test set to 0.0265 on average, MAPE from 0.8373 to 0.8597, and WIS$_{0.1-0.9}$ from 0.6717 to 0.7001.
The mean robustness score is 0.8884, and the radar plot and card text flag noise as a scenario where the newer model weakens despite clean MSE improvement.
The MLflow-backed bundle can support dashboards and be regenerated after retraining or scheduled data reviews.

\section{Discussion and Limitations}

IARC-TS connects governance and standards to measurement and reporting for soft sensing, monitoring, prognostics, anomaly detection, and quality estimation.
It can document statistical or transformer forecasters when they expose comparable predictions, intervals, and temporal validation records.
Closed-loop reinforcement learning and direct-control policies still need system-level validation because IARC-TS evaluates replayed or simulated scenarios, not closed-loop stability.

Compared with general-purpose model cards, IARC-TS is narrower but more prescriptive, fixing temporal splits, ODD coverage, scenario robustness, uncertainty calibration, and monitoring actions.
It separates model-side evidence from the plant-level safety case while keeping a lightweight interface for industrial projects.

The IndPenSim study is synthetic, so it cannot validate plant-specific instrumentation, real fermentation variability, or operator response.
Explainability, human-machine interaction, cybersecurity, hardware documentation, plant-level risk management, and data governance remain outside IARC-TS, but should link through the same model ID, dataset version, ODD, and scenario catalog.

% \section{Conclusion}

% A conclusion section is not required. Although a conclusion may review
% the main points of the paper, do not replicate the abstract as the
% conclusion. A conclusion might elaborate on the importance of the work
% or suggest applications and extensions.

% \begin{ack}
% Place acknowledgments here.
% \end{ack}

\section*{DECLARATION OF GENERATIVE AI AND AI-ASSISTED TECHNOLOGIES IN THE WRITING PROCESS}
During the preparation of this work, the author(s) used GPT-5 by OpenAI for editorial feedback. After using this tool/service, the author(s) reviewed and edited the content as needed and take(s) full responsibility for the content of the publication.

\bibliography{references}             % bib file to produce the bibliography

@inproceedings{windmann_artificial_2024,
    title = {Artificial {Intelligence} in {Industry} 4.0: {A} {Review} of {Integration} {Challenges} for {Industrial} {Systems}},
    copyright = {All rights reserved},
    shorttitle = {Artificial {Intelligence} in {Industry} 4.0},
    doi = {10.1109/INDIN58382.2024.10774364},
    urldate = {2025-01-10},
    booktitle = {2024 {IEEE} 22nd {International} {Conference} on {Industrial} {Informatics} ({INDIN})},
    author = {Windmann, Alexander and Wittenberg, Philipp and Schieseck, Marvin and Niggemann, Oliver},
    month = aug,
    year = {2024},
    pages = {1--8},
}

@misc{RegulationEU20242024,
  title = {Regulation ({EU}) 2024/1689 of the {{European Parliament}} and of the {{Council}} of 13 {{June}} 2024 ({Artificial Intelligence Act})},
  key = {EU AI Act},
  year = {2024},
  howpublished = {OJ L, 2024/1689, 12.7.2024},
  url = {http://data.europa.eu/eli/reg/2024/1689/oj},
  note = {Official EUR-Lex text},
  langid = {english}
}

@techreport{tabassiArtificialIntelligenceRisk2023,
  title = {Artificial {{Intelligence Risk Management Framework}} ({{AI RMF}} 1.0)},
  author = {Tabassi, Elham},
  year = {2023},
  month = jan,
  number = {NIST AI 100-1},
  pages = {NIST AI 100-1},
  address = {Gaithersburg, MD},
  institution = {{National Institute of Standards and Technology (U.S.)}},
  doi = {10.6028/NIST.AI.100-1},
  urldate = {2025-10-14},
  langid = {english}
}

@misc{ISOIECTR2021,
  title = {{{ISO}}/{{IEC TR}} 24029-1:2021 --- {{Artificial Intelligence}} ({{AI}}) --- {{Assessment}} of the Robustness of Neural Networks --- {{Part}} 1: {{Overview}}},
  shorttitle = {{{ISO}}/{{IEC TR}} 24029-1},
  key = {{{ISO}}/{{IEC TR}} 24029-1},
  year = {2021},
  urldate = {2024-01-12},
  langid = {english}
}

@misc{DINSPEC9200122020,
  title = {{{DIN SPEC}} 92001-2 - {{Artificial Intelligence}} - {{Life Cycle Processes}} and {{Quality Requirements}} - {{Part}} 2: {{Robustness}}},
  shorttitle = {{{DIN SPEC}} 92001-2},
  key = {{{DIN SPEC}} 92001-2},
  year = {2020},
  month = dec,
  publisher = {Beuth Verlag GmbH},
  doi = {10.31030/3205018},
  urldate = {2023-11-30}
}

@inproceedings{mitchellModelCardsModel2019,
  title = {Model {{Cards}} for {{Model Reporting}}},
  booktitle = {Proceedings of the {{Conference}} on {{Fairness}}, {{Accountability}}, and {{Transparency}}},
  author = {Mitchell, Margaret and Wu, Simone and Zaldivar, Andrew and Barnes, Parker and Vasserman, Lucy and Hutchinson, Ben and Spitzer, Elena and Raji, Inioluwa Deborah and Gebru, Timnit},
  year = {2019},
  month = jan,
  series = {{{FAT}}* '19},
  pages = {220--229},
  publisher = {Association for Computing Machinery},
  address = {New York, NY, USA},
  doi = {10.1145/3287560.3287596},
  urldate = {2025-02-05},
  isbn = {978-1-4503-6125-5}
}

@inproceedings{brajovicMergingEURegulationModel2023,
  title = {Merging ({{EU}})-{{Regulation}} and {{Model Reporting}}},
  booktitle = {{{NeurIPS}} 2023 {{Workshop}} on {{Regulatable ML}}},
  author = {Brajovic, Danilo and G{\"o}bels, Vincent Philipp and Kutz, Janika and Huber, Marco},
  year = {2023},
  month = dec,
  urldate = {2024-11-28},
  langid = {english}
}

@article{arnoldFactSheetsIncreasingTrust2019,
  title = {{{FactSheets}}: {{Increasing}} Trust in {{AI}} Services through Supplier's Declarations of Conformity},
  shorttitle = {{{FactSheets}}},
  author = {Arnold, M. and Bellamy, R. K. E. and Hind, M. and Houde, S. and Mehta, S. and Mojsilovi{\'c}, A. and Nair, R. and Ramamurthy, K. Natesan and Olteanu, A. and Piorkowski, D. and Reimer, D. and Richards, J. and Tsay, J. and Varshney, K. R.},
  year = {2019},
  month = jul,
  journal = {IBM Journal of Research and Development},
  volume = {63},
  number = {4/5},
  pages = {6:1-6:13},
  issn = {0018-8646},
  doi = {10.1147/JRD.2019.2942288},
  urldate = {2025-10-15}
}

@article{perez-cerrolazaArtificialIntelligenceSafetyCritical2024,
  title = {Artificial {{Intelligence}} for {{Safety-Critical Systems}} in {{Industrial}} and {{Transportation Domains}}: {{A Survey}}},
  shorttitle = {Artificial {{Intelligence}} for {{Safety-Critical Systems}} in {{Industrial}} and {{Transportation Domains}}},
  author = {{Perez-Cerrolaza}, Jon and Abella, Jaume and Borg, Markus and Donzella, Carlo and Cerquides, Jes{\'u}s and Cazorla, Francisco J. and Englund, Cristofer and Tauber, Markus and Nikolakopoulos, George and Flores, Jose Luis},
  year = {2024},
  month = apr,
  journal = {ACM Comput. Surv.},
  volume = {56},
  number = {7},
  pages = {176:1--176:40},
  issn = {0360-0300},
  doi = {10.1145/3626314},
  urldate = {2025-06-13}
}

@misc{ISOIEC2402922023,
  title = {{{ISO}}/{{IEC}} 24029-2:2023 --- {{Artificial}} Intelligence ({{AI}}) --- {{Assessment}} of the Robustness of Neural Networks --- {{Part}} 2: {{Methodology}} for the Use of Formal Methods},
  key = {ISO/IEC 24029-2},
  shorttitle = {{{ISO}}/{{IEC}} 24029-2},
  year = {2023},
  urldate = {2024-01-12},
  langid = {english}
}

@techreport{easaEASAArtificialIntelligence2024,
  title = {{{EASA Artificial Intelligence}} ({{AI}}) {{Concept Paper Issue}} 2: {{Guidance}} for {{Level}} 1\&2 Machine Learning Applications},
  author = {EASA},
  institution = {European Union Aviation Safety Agency},
  year = {2024},
  month = apr,
  urldate = {2025-05-27},
  langid = {english}
}

@article{lavinTechnologyReadinessLevels2022,
  title = {Technology Readiness Levels for Machine Learning Systems},
  author = {Lavin, Alexander and {Gilligan-Lee}, Ciar{\'a}n M. and Visnjic, Alessya and Ganju, Siddha and Newman, Dava and Ganguly, Sujoy and Lange, Danny and Baydin, At{\'i}l{\'i}m G{\"u}ne{\c s} and Sharma, Amit and Gibson, Adam and Zheng, Stephan and Xing, Eric P. and Mattmann, Chris and Parr, James and Gal, Yarin},
  year = 2022,
  month = oct,
  journal = {Nature Communications},
  volume = {13},
  number = {1},
  pages = {6039},
  issn = {2041-1723},
  doi = {10.1038/s41467-022-33128-9},
  urldate = {2023-08-02},
  langid = {english}
}

@inproceedings{windmannQuantifyingRobustnessBenchmarking2025,
  title = {Quantifying {{Robustness}}: {{A Benchmarking Framework}} for {{Deep Learning Forecasting}} in {{Cyber-Physical Systems}}},
  shorttitle = {Quantifying {{Robustness}}},
  booktitle = {2025 {{IEEE}} 30th {{International Conference}} on {{Emerging Technologies}} and {{Factory Automation}} ({{ETFA}})},
  author = {Windmann, Alexander and Steude, Henrik and Boschmann, Daniel and Niggemann, Oliver},
  year = 2025,
  month = sep,
  pages = {1--8},
  issn = {1946-0759},
  doi = {10.1109/ETFA65518.2025.11205527},
  urldate = {2025-10-22}
}

@article{gawlikowski2022surveyuncertaintydeepneural,
  title = {A Survey of Uncertainty in Deep Neural Networks},
  author = {Gawlikowski, Jakob and Tassi, Cedrique Rovile Njieutcheu and Ali, Mohsin and Lee, Jongseok and Humt, Matthias and Feng, Jianxiang and Kruspe, Anna and Triebel, Rudolph and Jung, Peter and Roscher, Ribana and Shahzad, Muhammad and Yang, Wen and Bamler, Richard and Zhu, Xiao Xiang},
  year = 2023,
  month = oct,
  journal = {Artificial Intelligence Review},
  volume = {56},
  number = {1},
  pages = {1513--1589},
  issn = {1573-7462},
  doi = {10.1007/s10462-023-10562-9},
  urldate = {2025-12-01},
  langid = {english}
}

@article{bracherEvaluatingEpidemicForecasts2021,
  title = {Evaluating epidemic forecasts in an interval format},
  author = {Bracher, Johannes and Ray, Evan L. and Gneiting, Tilmann and Reich, Nicholas G.},
  editor = {Pitzer, Virginia E.},
  year = 2021,
  month = feb,
  journal = {PLOS Computational Biology},
  volume = {17},
  number = {2},
  pages = {e1008618},
  issn = {1553-7358},
  doi = {10.1371/journal.pcbi.1008618},
  urldate = {2026-05-10},
  langid = {english}
}

@article{gneitingStrictlyProperScoring2007,
  title = {Strictly {{Proper Scoring Rules}}, {{Prediction}}, and {{Estimation}}},
  author = {Gneiting, Tilmann and Raftery, Adrian E},
  year = 2007,
  month = mar,
  journal = {Journal of the American Statistical Association},
  volume = {102},
  number = {477},
  pages = {359--378},
  issn = {0162-1459},
  doi = {10.1198/016214506000001437},
  urldate = {2026-05-10},
  langid = {english}
}

@misc{ISOIEC420012023,
  title = {{{ISO}}/{{IEC}} 42001:2023 - {{Information}} Technology --- {{Artificial}} Intelligence --- {{Management}} System},
  key = {{{ISO}}/{{IEC}} 42001},
  year = 2023,
  urldate = {2025-05-16},
  langid = {english}
}

@article{GOLDRICK2019106471,
title = {Modern day monitoring and control challenges outlined on an industrial-scale benchmark fermentation process},
journal = {Computers \& Chemical Engineering},
volume = {130},
pages = {106471},
year = {2019},
issn = {0098-1354},
doi = {10.1016/j.compchemeng.2019.05.037},
author = {Stephen Goldrick and Carlos A. Duran-Villalobos and Karolis Jankauskas and David Lovett and Suzanne S. Farid and Barry Lennox}
}

@misc{GMPAIRegulations,
  title = {{{Stakeholders' Consultation}} on {{EudraLex}} {{Volume}} 4 - {{Good Manufacturing Practice Guidelines}}: {{Chapter}} 4, {{Annex}} 11 and {{New Annex}} 22},
  key = {EudraLex GMP Annex 22 consultation},
  year = {2025},
  howpublished = {European Commission Public Health consultation page},
  note = {Opened 7 July 2025, closed 7 October 2025},
  langid = {english}
}

@inproceedings{hutchinsonAccountabilityMachineLearning2021,
  title = {Towards {{Accountability}} for {{Machine Learning Datasets}}: {{Practices}} from {{Software Engineering}} and {{Infrastructure}}},
  shorttitle = {Towards {{Accountability}} for {{Machine Learning Datasets}}},
  booktitle = {Proceedings of the 2021 {{ACM Conference}} on {{Fairness}}, {{Accountability}}, and {{Transparency}}},
  author = {Hutchinson, Ben and Smart, Andrew and Hanna, Alex and Denton, Remi and Greer, Christina and Kjartansson, Oddur and Barnes, Parker and Mitchell, Margaret},
  year = 2021,
  month = mar,
  pages = {560--575},
  publisher = {ACM},
  address = {Virtual Event Canada},
  doi = {10.1145/3442188.3445918},
  urldate = {2025-11-19},
  isbn = {978-1-4503-8309-7},
  langid = {english}
}

@incollection{niggemannMachineLearningCyberPhysical2023,
  title = {Machine {{Learning}} for {{Cyber-Physical Systems}}},
  booktitle = {Digital {{Transformation}}: {{Core Technologies}} and {{Emerging Topics}} from a {{Computer Science Perspective}}},
  author = {Niggemann, Oliver and Zimmering, Bernd and Steude, Henrik and Augustin, Jan Lukas and Windmann, Alexander and Multaheb, Samim},
  editor = {{Vogel-Heuser}, Birgit and Wimmer, Manuel},
  year = 2023,
  pages = {415--446},
  publisher = {Springer},
  address = {Berlin, Heidelberg},
  doi = {10.1007/978-3-662-65004-2\_17},
  urldate = {2023-02-02},
  copyright = {All rights reserved},
  isbn = {978-3-662-65004-2},
  langid = {english}
}

@article{dobbelaereMachineLearningChemical2021,
  title = {Machine {{Learning}} in {{Chemical Engineering}}: {{Strengths}}, {{Weaknesses}}, {{Opportunities}}, and {{Threats}}},
  shorttitle = {Machine {{Learning}} in {{Chemical Engineering}}},
  author = {Dobbelaere, Maarten R. and Plehiers, Pieter P. and Van De Vijver, Ruben and Stevens, Christian V. and Van Geem, Kevin M.},
  year = 2021,
  month = sep,
  journal = {Engineering},
  volume = {7},
  number = {9},
  pages = {1201--1211},
  issn = {20958099},
  doi = {10.1016/j.eng.2021.03.019},
  urldate = {2025-11-27},
  langid = {english}
}

@article{diaz-rodriguezConnectingDotsTrustworthy2023,
  title = {Connecting the Dots in Trustworthy {{Artificial Intelligence}}: {{From AI}} Principles, Ethics, and Key Requirements to Responsible {{AI}} Systems and Regulation},
  shorttitle = {Connecting the Dots in Trustworthy {{Artificial Intelligence}}},
  author = {{D{\'i}az-Rodr{\'i}guez}, Natalia and Del Ser, Javier and Coeckelbergh, Mark and {L{\'o}pez de Prado}, Marcos and {Herrera-Viedma}, Enrique and Herrera, Francisco},
  year = 2023,
  month = nov,
  journal = {Information Fusion},
  volume = {99},
  pages = {101896},
  issn = {1566-2535},
  doi = {10.1016/j.inffus.2023.101896},
  urldate = {2025-03-14}
}

@article{ashmoreAssuringMachineLearning2021,
  title = {Assuring the {{Machine Learning Lifecycle}}: {{Desiderata}}, {{Methods}}, and {{Challenges}}},
  shorttitle = {Assuring the {{Machine Learning Lifecycle}}},
  author = {Ashmore, Rob and Calinescu, Radu and Paterson, Colin},
  year = 2021,
  month = may,
  journal = {ACM Computing Surveys},
  volume = {54},
  number = {5},
  pages = {1--39},
  issn = {0360-0300, 1557-7341},
  doi = {10.1145/3453444},
  urldate = {2023-08-02},
  langid = {english}
}

@misc{ISOIEC238942023,
  title = {{{ISO}}/{{IEC}} 23894:2023 - {{Information}} Technology --- {{Artificial}} Intelligence --- {{Guidance}} on Risk Management},
  key = {{{ISO}}/{{IEC}} 23894},
  year = 2023,
  urldate = {2024-02-28},
  langid = {english}
}

@inproceedings{vranjesDesignPrinciplesFalsifiable2024,
  title = {Design {{Principles}} for {{Falsifiable}}, {{Replicable}} and {{Reproducible Empirical Machine Learning Research}}},
  booktitle = {35th {{International Conference}} on {{Principles}} of {{Diagnosis}} and {{Resilient Systems}} ({{DX}} 2024)},
  author = {Vranje{\v s}, Daniel and Ehrhardt, Jonas and Heesch, Ren{\'e} and Moddemann, Lukas and Steude, Henrik Sebastian and Niggemann, Oliver},
  year = 2024,
  doi = {10.4230/OASIcs.DX.2024.7},
  urldate = {2025-12-02}
}

@article{WebbCharacterizingconceptdrift2016,
   title={Characterizing concept drift},
   volume={30},
   ISSN={1573-756X},
   DOI={10.1007/s10618-015-0448-4},
   number={4},
   journal={Data Mining and Knowledge Discovery},
   publisher={Springer Science and Business Media LLC},
   author={Webb, Geoffrey I. and Hyde, Roy and Cao, Hong and Nguyen, Hai Long and Petitjean, Francois},
   year={2016},
   month=apr, pages={964–994} }
                                                     
% \appendix
% \section{A summary of Latin grammar}    % Each appendix must have a short title.
\end{document}